# Epitaxial Ni/Cu Superlattice Nanowires with Atomically Sharp Interfaces for Spin Transport

Janez Zavašnik [1,2], Sama Derakshan-Nejad [3], Maryam Ghaffari [3], Amir Hassan Montazer [4], Mohammad Reza Mardaneh [3], Mohammad Almasi Kashi [3], Alexandre Nominé [1,5], Stephane Mangin [5,6], Uroš Cvelbar[1,7]

[1]Jožef Stefan Institute, Jamova cesta 39, 1000 Ljubljana, Slovenia
[2]Max-Planck-Institut für Nachhaltige Materialien; Max-Planck-Straße 1, 40237 Düsseldorf, Germany
[3]Institute of Nanoscience and Nanotechnology, Department of Physics, University of Kashan, Kashan, Qotb-e Ravandi Blvd, Iran
[4]Department of Medical Equipment Technology Engineering, Al-Hadba University, Mosul, Nineveh Governorate, 41002, Iraq
[5]Institut Jean Lamour – CNRS – Université de Lorraine, Campus ARTEM, 2 allée André Guinier, 54011 Nancy, France
[6]Institut Universitaire de France (IUF), 1 rue Descartes, 75231 Paris cedex 05, France
[7]Jožef Stefan International Postgraduate School, Jamova c. 39, 1000 Ljubljana, Slovenia

**Abstract**

The importance of microstructure increases when decreasing the size of an object to the nanoscale, along with the complexity of controlling it. For instance, it is particularly complicated to create nano-object with controlled interfaces. Therefore, progressing towards 1D epitaxial nanostructures poses a challenge, and realization of their full potential is linked to technological issues of achieving large-scale, precise atom stacking of two or more different chemical elements. Achieving such coherent, epitaxial interfaces is a key step toward enabling spintronic phenomena in 1D objects, by minimizing interface scattering and strain-driven defects. Our results demonstrate a successful realization of controlled nanoscale heteroepitaxy in one-dimensional single-crystal structures. We fabricated nanowires composed of alternating magnetic (nickel) and non-magnetic, highly conductive (copper) segments. This periodic stacking modulates electron transport under magnetic stimuli. The epitaxial precision achieved eliminates detrimental electron scattering that has historically limited the magnetotransport properties of such 1D structures and hindered their development. Such materials are crucial for further advancements in the miniaturisation of nanosensors, actuators, and next-generation 3D spintronic devices.

## 1. Introduction

The properties of materials at reduced dimensions often significantly differ from their bulk macroscopic counterparts. Similarly, crystal-structure-related phenomena, such as efficient charge and excitons transfer, become more pronounced and are easier to measure, interpret, and exploit when manifested in single crystals [1]. The fabrication of single-crystal one-dimensional (1D) metal nanowires with well-developed morphology has long been a pursuit in materials science, as they represent ideal building blocks for nanoscale electronics, sensors, actuators, electrodes and antennae. A variety of synthesis methods have been developed for their production, including vapour phase deposition [2], thermal annealing [3], and electrodeposition [4], all of which are well-characterised and widely explored.

Unlike single-phase nanowires, multi-layered nanowires composed of two or more metals are a special case due to their spatial modulation, multiple functionalities, and enhanced physical properties [5–7]. Up to now, such multi-material layered 1D nanowires were polycrystalline, and the controlled synthesis of heterogeneous and at the same time single-crystal nanowires proved to be challenging, as they are neither thermodynamically nor kinetically favourable [8]. When successfully realised, metals with significantly different physical properties can be joined within a single crystal. When alternating magnetic and non-magnetic domains, such segment-modulated multi-layered (barcode) nanowires can enable the design of tuneable microwave absorbers, high-sensitivity magnetic sensors, biomimetics actuators, and advanced 3D spintronic components such as racetrack memories with ultrahigh storage densities, operating by spin transfer and spin-orbit torque mechanisms.

The bulk multi-component epitaxial systems of highly ordered atomic arrangements were realised for the first time by molecular-beam epitaxy (MBE) and are nowadays routinely used for large-scale thin-film deposition of single crystals, among other also magnetic materials [9–11]. Nevertheless, the further need for miniaturisation demands the transition from two-dimensional thin films to one-dimensional epitaxial nanosystems [12,13]. The discovery of giant magnetoresistance (GMR) in metallic multilayers initially sparked great interest in 1D nanostructures. However, despite early enthusiasm, the performance of nanowire-based devices failed to meet expectations, largely due to their uncontrolled polycrystalline microstructure and incoherent interfaces, which limited spin transport and compromised reproducibility. Up to now, the electrodeposition synthesis of segmented bilayer polycrystalline nanowires showing interesting magnetic properties has been demonstrated in several systems (Co/Cu, Ni-Fe/Cu and Ni/Cu), showing successful nanoscale alteration of the magnetic and non-magnetic stacks [14–17]. But for the successful interplay between the magnetic field and conductivity, the simple thickness control of magnetic and non-magnetic domains is not enough. Despite extensive research, the electro-deposited multi-layer nanowires suffer the same challenge as the initial steps of MBE thin-film development: the lack of the epitaxial intergrowth of the magnetic and non-magnetic spacer domains (segments), which is needed to prevent electron scattering on the interface between both counterparts. Although the paradigm of material being simultaneously heterogeneous and single-crystal is contradictive, it can be realised by the crystal epitaxy in materials having homologous crystal structures.

For our case study of modulated magnetic superlattice nanowires, we choose Cu and Ni as they appear to be a perfect match for the heteroepitaxial, periodically layered nanowire demonstration system: copper is well-known for its excellent conductivity while at the same time being practically magnetically inert. On the other hand, ferromagnetic nickel is a suitable candidate for the magnetically active spin valve component of the giant magnetoresistance (GMR) stack, with total magnetization of 0.71 µB/f.u. Both Cu and Ni crystallise in the face-centred cubic (fcc) structure, have similar atomic radii, electronegativities, and valences, and are, therefore, good candidates for the design and manufacture of the epitaxial superstructures.

In our experiment, we analysed Ni/Cu nanowires prepared by template-pulsed electrodeposition (TPED) into an anode aluminium oxide (AAO) template, a synthesis method widely used due to its simplicity, scalability, and cost-effectiveness [18,19]. We achieved precise Ni-Cu sequential stacking with atomically sharp interfaces, enabled by the kinetic control of heteronucleation under near-equilibrium conditions. Multiple atomic stacking orders at the Ni-Cu interfaces were confirmed via XRD, scanning, and transmission electron microscopy. The resulting heterostructures exhibited a structurally coherent superlattice, confirming that pulsed electrodeposition is a viable route for producing nanostructured, compositionally modulated, epitaxial single-crystal materials.

## 2. Results and Discussion

*2.1 The heteroepitaxial growth of Ni/Cu nanowires was achieved via pulsed electrodeposition into porous anodic alumina membranes.* Morphological analysis of the as-deposited NW arrays, still embedded in the porous alumina membrane, revealed uniform nanowires approximately 55 ± 5nm in diameter, with an inter-wire spacing of ~105 nm. The pore-filling efficiency exceeded 80%, with unfilled pores primarily localized at the alumina grain boundaries (**Figure 1a**). The nanowires have an average length of ~3.5 μm, populating the lower part of the 10 μm thick AAO template channels (**Figure 1b**). X-ray diffraction (XRD) served as a rapid screening tool for phase and interface quality, enabling real-time feedback during deposition optimization. The diffraction spectra of embedded nanowires showed distinct peaks corresponding to pure fcc Ni and Cu, with no detectable Ni-Cu alloy formation (**Figure 1c**). The satellite reflections around the Ni(111) peak at m(-1) at 74.75° 2θ and m(+1) at 77.11°, which arise from periodic Ni–Cu segment alternation, confirm the formation of a coherent crystallographic superlattice.

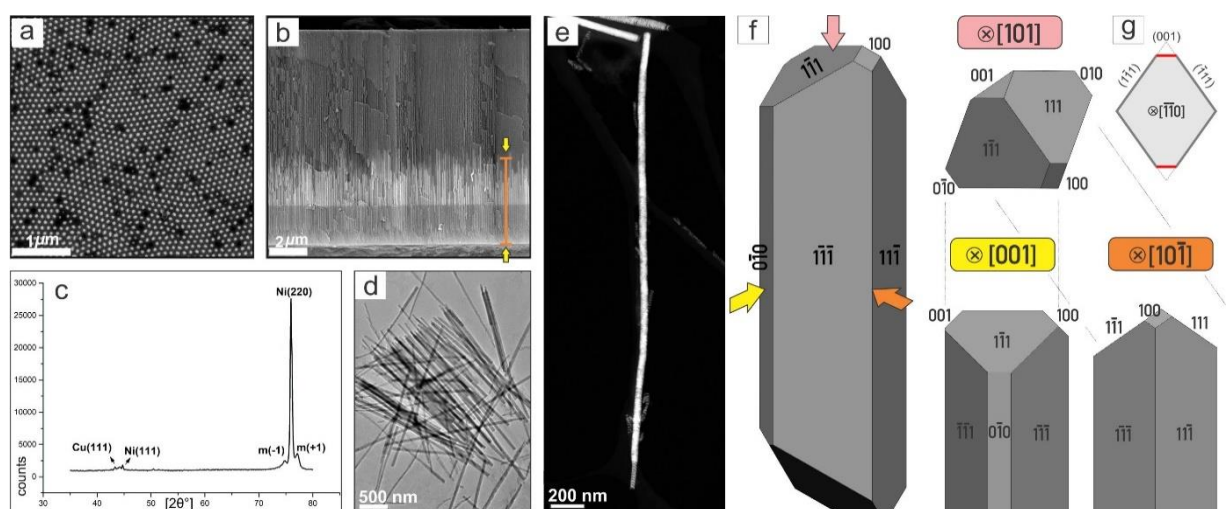

**Figure 1**. Crystallography and morphology of the nanowires. (a) top-view and (b) cross-sectional SEM-BEI images of Ni/Cu multi-layered segmented NW arrays still embedded in the porous alumina membrane (NWs filling the pores are marked by arrows). (c) XRD of Ni-Cu multilayer NWs showing satellite peaks originating from periodic alterations of the Ni and Cu layers, forming superlattice. (d) TEM overview micrograph of Ni/Cu NWs after release from the template and deposited onto the TEM support grid; NWs tend to form bundles due to the magnetic attraction. (e) Physical appearance and morphology of the isolated NW; HAADF-STEM micrograph. (f) Reconstructed morphology and shape of NW with marked crystallographic data. (g) Equilibrium cross-sectional shape of the Ni/Cu [110] NW calculated by Wulff construction.

After the release from the alumina template, the nanowires were washed, centrifuged in EtOH, and transferred onto TEM amorphous carbon support grids. Due to their residual magnetic properties, the nanowires tend to aggregate into aligned bundles (**Figure 1d**). High-angle annular dark-field scanning transmission electron microscopy (HAADF-STEM) imaging revealed a short initial stem-like growth segment (~200 nm) followed by periodic, compositionally alternating domains of similar morphology (**Figure 1e**). The nanowire shape was reconstructed from the HAADF-STEM tilt-series (see Supplementary Information File for details). The individual NW has a wedge-shaped faceted geometry, composed of <111> and <100> facets resulting in a truncated rhombic cross-section (**Figure 1f** and **1g**). This faceting contrasts with the cylindrical morphology typically produced by electrodeposition in AAO templates and was previously observed only in wires with diameters below 50 nm [20,21], suggesting a strong influence of crystallographic epitaxy on morphological anisotropy.

*2.2 The compositional and crystallographic structure of multilayer Ni/Cu nanowires was confirmed by EDXS mapping, high-resolution TEM, and selected area electron diffraction (SAED).* All nanowires exhibited a consistent [110] growth direction, with their top ends bounded by {111} facets forming a wedge-shaped termination (**Figure 2a**). Despite the atomic-resolution imaging indicating structural continuity (i.e. coherent, single-crystal structure), the Energy Dispersive X-ray Spectroscopy (EDXS) elemental mapping revealed clear compositional segmentation: alternating Ni-rich and Cu-rich domains forming a periodic fishbone contrast (**Figure 2b**, and **2c**). High-resolution TEM confirmed that both Ni and Cu segments share a common crystallographic orientation, indicating coherent epitaxial alignment across the interfaces (**Figure 2d** and **2e**). The electron diffraction pattern further confirm the $(110)[111]_{Cu}||(110)[111]_{Ni}$ epitaxial relationship (**Figure 2f**). Interface sharpness and segment thickness are uniform and maintained along the full nanowire length, comprising approximately 400 bilayers. For example, the representative NW shown in **Figure 2a-e** had alternating 10 nm Ni and 5 nm Cu layers, with other stacking geometries (e.g., 10/10) reproducibly achieved by tuning the deposition parameters (**Figure 2g**).

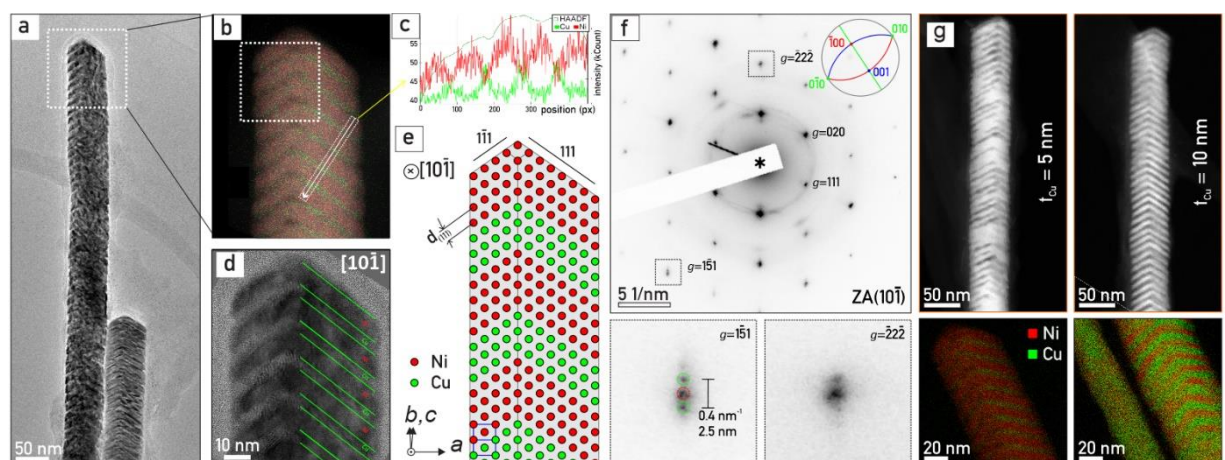

**Figure 2.** Compositional and crystallographic structure characterisation of multi-layered superlattice NWs. (a) TEM micrograph of Ni/Cu NW in [01-1] zone axis showing characteristic (110)-pointed fishbone-shaped segmentation. (b) STEM-EDXS elemental mapping (red = Ni, green = Cu) with (c) corresponding intensity profile. (d) HR-TEM micrograph of single-crystal epitaxial alteration of 10 nm Ni and 5 nm Cu segments. (e) Schematic diagram of atomic arrangements and crystal planes; {111} is a shared plane between Ni and Cu segments (not up to scale). (f) SAED pattern of a 10Ni/5Cu heterostructure. Diffraction peaks marked with dashed squares are magnified below to show details of the Bragg and satellite peaks. The Bragg peaks of Ni and Cu are marked with red and green circles, respectively, with the corresponding Miller indices shown. The satellite peaks indicate periodic modulation in the heterostructure. (g) An example of successful modulation of Cu-segment thickness; $t_{Cu}$ = 5 nm and $t_{Cu}$ = 10 nm, with corresponding EDXS maps (red = Ni, green = Cu).

*2.3 The crystal structure of the Ni/Cu nanowires was characterised by electron diffraction experiments (SAED) in transmission electron microscopy (TEM).* The SAED pattern from a representative nanowire composed of 10 nm Ni and 5 nm Cu layers (**Figure 2f**) shows two distinct sets of Bragg reflections corresponding to the Ni and Cu phases. Notably, additional satellite peaks appeared around the main reflections, indicating periodic lattice modulation in the multilayered structure (**Figure 2f**). The presence of the satellite peaks, under the kinematical approximation of electron diffraction, is a hallmark of well-ordered superlattices, i.e. the periodic lattice modulation. Detailed crystallographic analysis showed that the interface between Ni and Cu layers follows the {111} planes, with a lattice mismatch of approx. 2.5 % ($a_{Cu}(111)/a_{Ni}(111)$ = 1.0329). This mismatch

results in the accumulation of one extra Ni atom every 40 atomic planes (**Figure 2 d** and **2e**). The resulting strain is accommodated by visible contrast ripples in the Cu segments that we attribute to compressive–tensile stress compensation at the interface (**Figure 3a**). Atomic-resolution imaging revealed that this mismatch is periodically terminated by edge dislocations on both sides of each Cu segment, separated at regular intervals of ~20 atoms (**Figure 3b**). Simulations of the Ni–Cu interface structure closely matched these experimental features, confirming the lattice registry and mismatch accommodation mechanisms (**Figure 3c**).

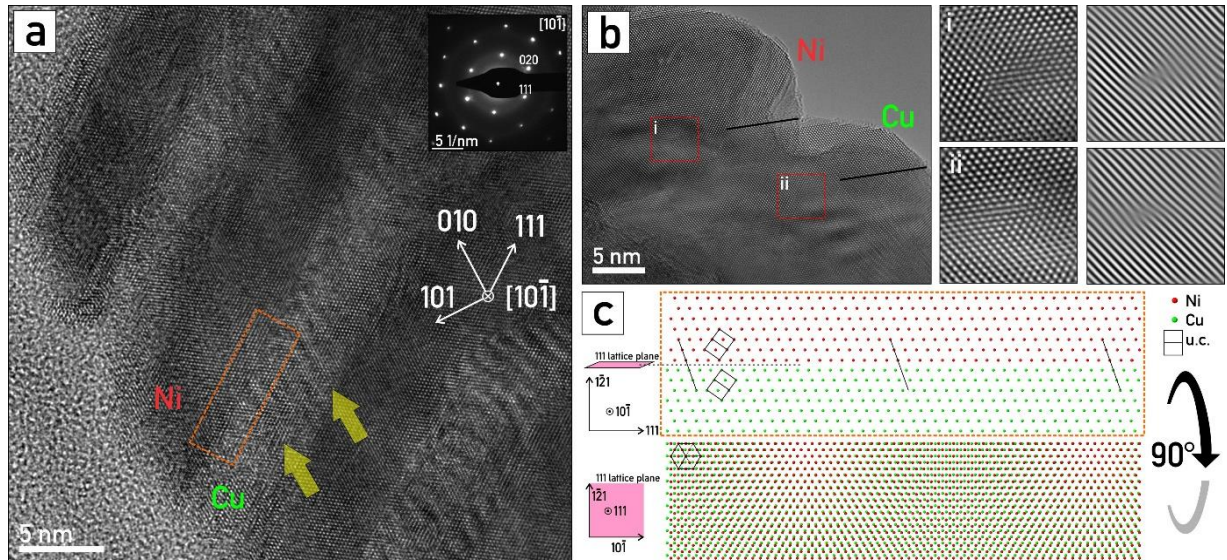

**Figure 3.** The epitaxial interface between Ni and Cu segments, and lattice mismatch compensation. (a) HR-TEM micrograph of the Ni-Cu interface detail, with SAED for plane identification. The yellow arrows mark the contrast variation present in the Cu segments due to the slight lattice mismatch between Ni and Cu. (b) Details of Ni-Cu interface, with the insets (i,ii) of TEM micrograph and inverse FFT of the detail of extra Ni atom layer. (c) Crystal structure simulation of Ni – Cu lattice mismatch and resulting Moiré pattern, in ZA[100], and rotated for 90 °, ZA(1-10). The 2.5 % lattice mismatch results in 1 extra Ni atom every 40 Cu atoms.

*2.4 The observation of GMR in Ni/Cu superlattice nanowires provides evidence that the heteroepitaxial interfaces between Ni and Cu layers are atomically sharp and structurally coherent.*
Earlier reports on GMR in polycrystalline Ni/Cu multi-layered nanowires fabricated via electrodeposition heavily rely on tuning of the Cu spacer thickness and optimizing interlayer magnetic coupling (RKKY) [15,17,22–24]. In contrast, our nanowires show epitaxial alignment across hundreds of bilayers, enabling a structurally coherent pathway for spin-polarized electron transport. The resulting measurable GMR effect points to the critical role of interface control for spin-polarized electron transport. The unpreceded precision of pulsed AC electrochemical deposition of segmented, epitaxial single-crystal Ni/Cu nanowires results in a well-aligned crystal structure across segments, which in turn enables spin-polarized electron transport. The Ni/Cu array show distinct magnetic anisotropy, with easy magnetization axis parallel to the wire growth direction. After their release from the AAO template, **Figure 4** shows the GMR (%) curve as a function of the magnetic field applied perpendicular to the aligned network of Ni/Cu multi-layered NWs. The analysed sensor is magnetically saturated at 500 Oe, presenting a maximum GMR effect of 2.4%. For comparison, magnetoresistance properties of a large 2D Ni/Cu multi-layered thin films reports maximum GMR values of up to 7% [25–27]; this discrepancy we attribute to inter-NW magnetic interactions and possible intra-NW magnetostatic coupling, with possible additional overall GMR decrease due to misaligned NWs during the measurements [17]. Nevertheless, the present study provides the first evidence of the fabrication of a high-sensitivity GMR sensor at room temperature based on the epitaxial superlattice Ni/Cu NWs. Such nano-scaled nanowires, due to their unique one-dimensional structure, can enable precise magnetic field detection and greater integration density in technological applications.

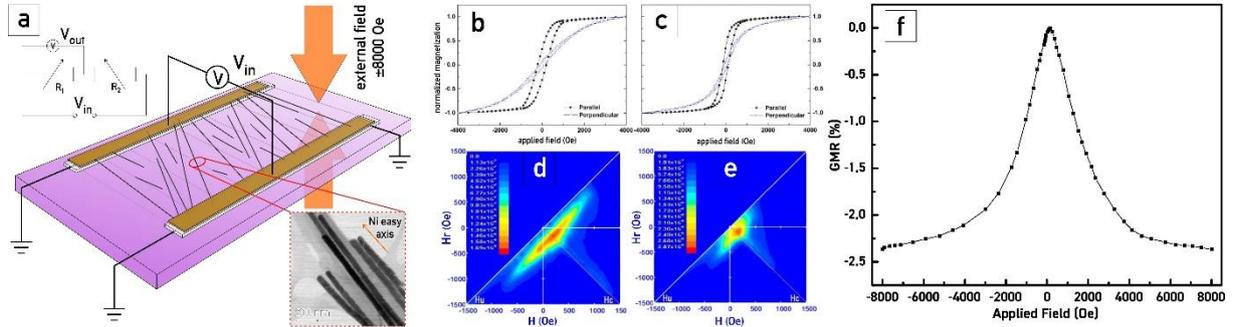

**Figure 4.** (a) Simplified GMR measurement circuit diagram and measurement scheme. Hysteresis loops and FORC diagrams of Ni/Cu multi-layered NWs for (b, d) $t_{Cu}$= 5 nm and (c, e) $t_{Cu}$= 10 nm. (f) Giant magnetoresistance (GMR) per cent curve as a function of a magnetic field applied perpendicular to an aligned network of Ni/Cu multi-layered NWs.

**Table 1.** Parallel ($\parallel$) and perpendicular ($\perp$) coercive field (Hc) and saturation magnetic field ($H_s$) extracted from hysteresis loop measurements of Ni/Cu multi-layered NWs with Cu segment thickness ($t_{Cu}$) of 5 and 10 nm. The respective FORC peak at $H_u$= 0 Oe ($H_c^{FORC}$) and magnetostatic interaction field distribute on ($\Delta H_u$) are also included.

| Ni/Cu multi-layered NWs | $H_c\parallel$ (Oe) | $H_c\perp$ (Oe) | $H_s\parallel$ (Oe) | $H_s\perp$ (Oe) | $H_c^{FORC}$ (Oe) | $\Delta H_u$ (Oe) |
|---|---|---|---|---|---|---|
| $t_{Cu}$= 5 nm | 207 | 111 | 1200 | 4000 | 180 | 478 |
| $t_{Cu}$= 10 nm | 125 | 70 | 950 | 3000 | 147 | 227 |

*2.5 The microstructure plays a crucial role in determining the final properties of the materials.*
Nanowires, with their unique one-dimensional structure, have distinct mechanical, electrical, and thermal properties that can be additionally finely tuned through microstructure control. In the view of these vast possibilities, the barcoded NWs have been extensively explored in the early 2000 for the GMR applications [28,29], with revived interest by further studies of spin-transfer torques [30], domain wall motion studies and definition of bits for a 3D race-track memory [31,32]. These approaches utilize alternation of hard and soft magnetic segments such as Co/Ni, or the combination of magnetic/nonmagnetic sections such as Co/Cu and Co/Au for magnetic nano-oscillators [33]. The examples on Co/Cu and FeNi/Cu provided GMR ratio in the range 14–20 % at $LN_2$ temperatures [34]. Most of the exemplary system were realised in planar stacks as the natural and easiest contacting geometry is current-in-plane (cip), while patterns with a vertical geometry suitable for the current-perpendicular-to-plane are challenging to achieve. The current perpendicular to plane (CCP) geometry has been applied later and reported 4.5 % GMR at room temperature [35]. Unfortunately, the lack of the control over the interface presented a major obstacle in production of repeatable benchmark nanostructures, and the efforts were eventually practically abandoned. The polycrystalline nature of the segments resulted in large scattering of the experimental results, indicating that the adjustments in grain size, crystal orientation, and even defect density can dramatically affect the conductivity, strength, and even the mechanical properties of the NW [34]. Therefore, the capability to engineer nanoscale, single-crystal nanowires, which is at the same time chemically heterogeneous, opens vast possibilities for creating tailored materials suited for specific applications, ranging from electronics to energy storage. While previous reports have demonstrated CPP-GMR in polycrystalline Ni/Cu nanowires with values reaching ~2.6 % through careful tuning of Cu spacer thickness and RKKY-driven interlayer coupling [17], our findings indicate that such magnetotransport effects can be achieved even with relatively modest segment tuning, provided that the interfaces are atomically sharp and crystallographically coherent. This highlights the distinct contribution of epitaxial registry to spin transport efficiency. Beyond microstructure manipulation, we expect that in the future the overall GMR response can be further improved by precise tailoring of the chemical composition by alloying [36], which has capacity to satisfy the simultaneous conditions of epitaxial growth and intrinsic magnetic properties.

## 3. Conclusions

The crystal structure and coherent interface in epitaxial multisegmented Ni/Cu superlattice nanowire arrays achieved by a pulsed AC electrochemical deposition into AAO templates were explored and interpreted, demonstrating that AAO synthesis is capable of precise deposition of chemical constituents on nano-level. Several different alternating Ni and Cu epitaxial layers of various thicknesses were explored, arranged in the superlattice. The ability to fabricate single-crystal nanowires with atomically sharp heterointerfaces between dissimilar metals was confirmed by GMR, serving as confirmation of the structural quality and coherent spin-transport pathways within the nanowires. The GMR effect is a consequence of atomic-scale heteroepitaxy, not merely layer thickness tuning or interfacial coupling strength. The importance of the result lies in the development of new spin device paradigms, from 2D to 1D materials, with possibility of precise magnetic field detection, design of microwave devices with tuneable filters and circulators, and greater integration density in technological applications.

## 4. Experimental Section/Methods

### 4.1. Experimental system

*4.1.1. Synthesis, pore widening and barrier layer thinning of porous alumina membrane.* High purity (99.999%, Chempure) aluminium foil was used as a starting material to synthesise the porous alumina membrane with hexagonally ordered nanopore arrays. The aluminium foil was washed in acetone in an ultrasonic bath and electropolished in a mixed solution of $HClO_4/C_2H_6O$ (1:4 in volume) at 3 °C at 20 V for 5 min. A two-step anodization process was used to form the porous alumina membrane. The first anodization step was performed in 0.3 M oxalic ($C_2H_2O_4$) acid solution under a potential of 40 V at 17 °C for 6 h. The oxide layer formed during the first step was etched in a mixed solution of 0.5 M phosphoric ($H_3PO_4$) and 0.2 M chromic ($CrH_2O_4$) acids at 60 °C for 6 h. The same conditions as those of the first step were applied to perform the second step, except that the anodization duration was 1 h, leading to highly ordered nanopores with 30 nm diameter, an interpore distance of 100 nm, [21] and a length of about 10 μm. The pore widening process of the alumina membrane was carried out by immersing it in 0.3 M $H_3PO_4$ acid solution at 30 °C for 15 min, increasing the pore diameter to about 50 nm. The alumina barrier layer was thinned by exponentially decreasing the anodization potential from 40 to 12 V in 0.3 M $C_2H_2O_4$ acid solution at 17 °C to facilitate the pulsed electrochemical deposition of NWs.

*4.1.2. Pulsed electrochemical deposition of Ni/Cu multi-layered NWs.* Ni/Cu multi-layered NWs with variable Cu segment thicknesses were electrochemically synthesised in the porous alumina membrane using a single bath, alternating current pulsed electrodeposition method: an electrochemical bath containing 0.6 M NiSO$_4$ · 6H$_2$O and 0.003 M CuSO$_4$ · 5H$_2$O together with 45 g/L H$_3$BO$_3$ was used at 30 °C and pH= 3, where the remaining aluminium of the membrane and a graphite roll served as the working and counter electrodes, respectively.[21] Sine waveforms with reduction/oxidation potentials of 15/11 V and 11/11 V along with reduction/oxidation times of 2.5/2.5 ms and 2.5/2.5 ms were created to fabricate Ni and Cu segments, respectively. The cathodic current density peak, off-time between pulses and number of Ni and Cu segments along the multi-layered NW length were kept constant to 105 mA/cm$^2$, 0 ms and 400, and 17 mA/cm$^2$, 200 ms and 400, respectively. Accordingly, 400 Ni/Cu double segments were fabricated along the NW length. While the Ni deposition pulse number remained constant to 50, the influence of Cu segment thickness ($t_{Cu}$) on the multi-layered NW properties was investigated by increasing the Cu deposition pulse number from 50 to 600.

**4.2. Material characterization**

To prepare the NW samples for analyses and characterisation, the remaining aluminium of the membrane sample was etched in a mixed saturated CuCl$_2$ solution for 5 min. The alumina barrier layer was then removed using 0.3 M H$_3$PO$_4$ acid solution for 20 min. Finally, the resulting sample was immersed in a 0.3 M NaOH solution for 90 min at room temperature, completely releasing the multi-layered NWs from the membrane.

Phase composition and crystal structure were investigated by X-ray diffraction (XRD, Philips X'Pert Pro; Cu-Kα radiation; λ=0.154 nm). Morphological properties of the porous alumina membrane and Ni/Cu multi-layered NW arrays were studied by field-emission scanning electron microscope (FESEM, MIRA3, TESCAN), operating at 15 kV. For nanoscale observations and analyses, we employed an aberration-corrected transmission electron microscope (S/TEM, Titan Themis, FEI) operating at 300 kV. The phase composition and crystal structure of the NWs were analysed by structure-dependent electron scattering by selected area electron diffraction (SAED). The phase composition was studied using high-angle annular dark-field (HAADF-STEM) microscopy. The chemical composition of the multi-layered NWs was characterised by energy-dispersive X-ray spectroscopy (EDS, Super X 4 SDD). Magnetic properties were studied using a vibrating sample magnetometer (VSM, MDK) at room temperature. Hysteresis loops were measured by applying a magnetic field H (± 4000 Oe) parallel and perpendicular to the NWs' axis. Moreover, first-order reversal curve (FORC) diagrams were obtained by saturating magnetisation M of multi-layered NWs using a positive $H_{Sat}$ applied parallel to the NWs' axis. The $H_{Sat}$ was then reduced to a reversal field $H_r$ while measuring $M(H, H_r)$. By sweeping $H_r$ back to $H_{Sat}$, a FORC was calculated. The FORC distribution is defined in Eq. 1 [22]:

$$\rho(H, H_r) = -\frac{1}{2}\frac{\partial^2 M(H, H_r)}{\partial H \partial H_r} \quad (1)$$

For convenience, coercive field ($H_c$) and interaction field ($H_u$) axes changed as follows: $H_c= (H–H_r)/2$ and $Hu= (H+H_r)/2$, representing a contour plot ranged from blue (minimum ρ) to red (maximum ρ) colour.

**4.3. GRM measurements**

*NW network preparation and GMR sensor fabrication.* A random network of Ni/Cu NWs (consisting of 400 Ni/Cu double segments) was initially prepared by complete dissolution of the aluminium substrate and a barrier layer of NW sample by saturated $CuCl_2$ and 0.3 M $H_3PO_4$ acid solution at 30 °C, respectively [23]. The NWs were released from the alumina membrane using 2 M NaOH solution for 1 h at room temperature, followed by washing with distilled water and ultrasonicated for 3 s. A drop of the solution containing NWs was placed on a clean glass substrate. An external magnetic field with a magnitude of 50 Oe was applied to the random NW network, aligning Ni/Cu NWs with respect to each other. The resulting parallel network of NWs was exposed to HCl vapour for 1 min to eliminate oxides and residues. The electrode for GMR measurements was fabricated by fixing and stuck two Cu contact pads on the NW network by Ag conductive paste. The NW network was annealed twice in an oven at 80 °C and 100 °C for 30 min to enhance the electrical conductivity. GMR measurements were carried out by applying a constant current between the Cu electrodes and varying magnetic field between ±8000 Oe perpendicular to the network plane.

**Acknowledgements**
This work was partially supported by the Slovenian Research and Innovation Agency (ARIS) program P1-0417 and project J2-440, and the interdisciplinary project LUE "MAT-PULSE", part of the French PIA project "Lorraine Université d'Excellence" reference ANR-15-IDEX-04-LUE. JZ acknowledges the support of Max-Planck- Gesellschaft via High performance materials Partner Group. The authors would like to acknowledge the financial support from the European Innovation Council project Pathfinder, No. 101046835.

**Data Availability Statement**
The data that support the plots within this paper and other findings of this study are available from the corresponding authors on reasonable request.

**Supporting Information**
Supporting Information is available from the Online version of the article or from the author.

# Supporting Information File

## 1. Nanowire deposition into Anode Aluminium Oxide (AAO) template

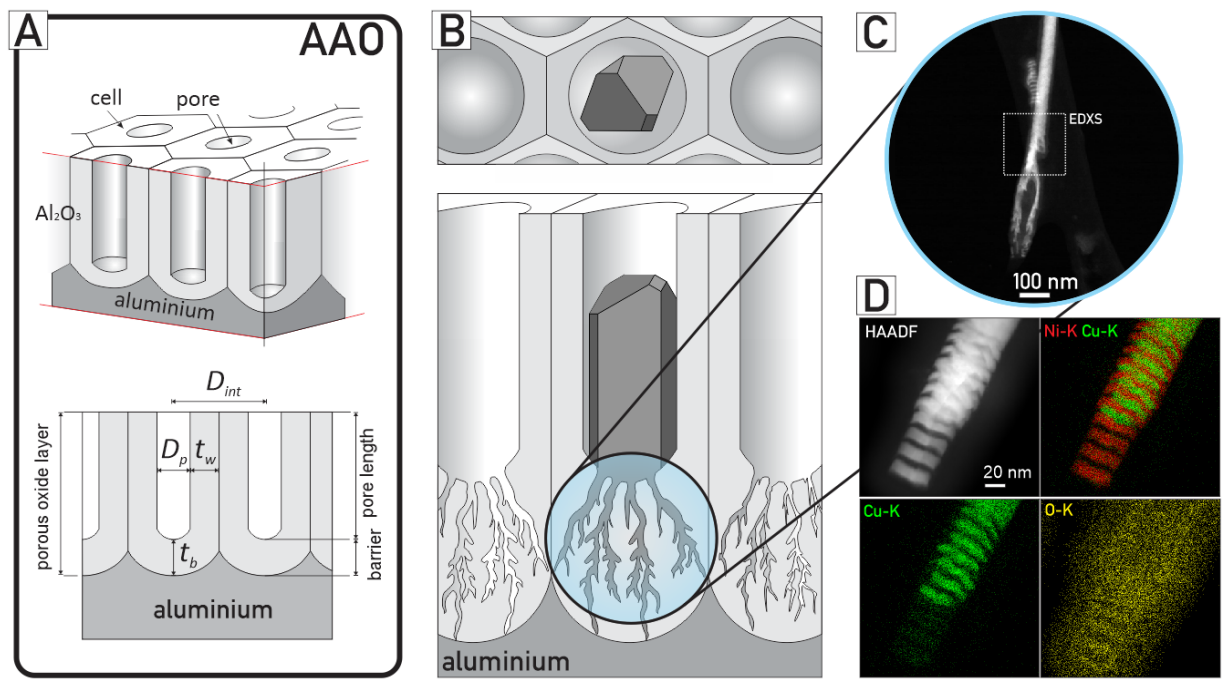

**Figure S1**: From nanopores array to nanowires array. (A) Schematic drawing of AAO, with marked main features: individual cell and pore in AAO developed on aluminium metal during anodization process, and main measurement quantities: pore diameter ($D_p$), interpore distance ($D_{int}$), AAO wall thickness ($t_w$), barrier layer thickness ($t_b$), and pore length and overall thickness of porous oxide layer. (B) Growth of faceted NW in AAO pore after barrier removal. (C) HAADF-STEM micrograph of the initial stage of NW growths in the near-pore dendritic section formed during the pore widening stage. (D) EDXS of the initial stages of the NW growth; here, the remaining oxygen in the solution is consumed, forming Cu-oxide in the first few segments and disk-like sections. After oxygen is spent, the sections uptake the wedge-like morphology, characteristic of epitaxial growth.

## 2. On the lattice parameters of multi-segmented Ni-Cu NWs as determined by XRD and TEM

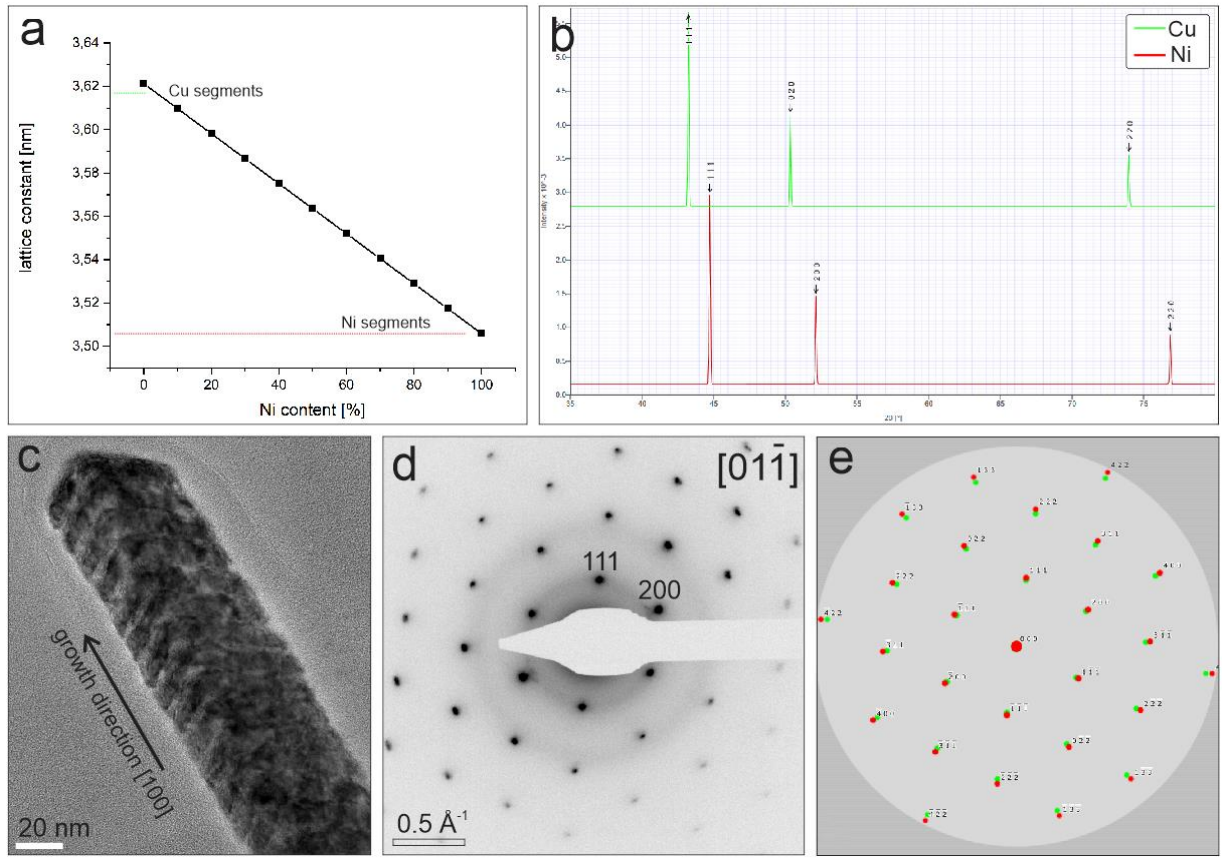

**Figure S2**: (a) The lattice constant as a function of the Ni-Cu ratio, with marked measured cell parameters from Ni (red) and Cu (green) segments, following Vegard's law. (b) Simulated XRD spectra for pure Ni and Cu phases, with evident peak shift due to the difference in the unit cell. (c) TEM micrograph of single Ni-Cu NW, with (d) corresponding SAED pattern, compared to (e) simulated SAED for pure Ni (red) and Cu (green) phases; the difference in electron diffraction scattering vector is small, making the distinguishing of individual peaks difficult.

**Supporting data Table S1**: Structure factors as calculated for pure Ni and Cu materials, compared to *d*-values calculated from experimental XRD spectra and from TEM analyses (measured from SAEDP and calculated from FFT) (bold = **measured**, italic = *calculated*). Reference data: cubic Fm-3m, Ni a= 3.50580 Å; Cu a = 3.62126 Å

| Ni(hkl) | *fcc* Ni ref [Å] | Ni ref (XRD) [°] | Ni exp (XRD) [°] | Ni exp (XRD) [Å] | Ni exp (TEM) [Å] |
|---|---|---|---|---|---|
| 111 | 2.02407 | 44.738 | **44.75** | 2.025 | **2.024** |
| 200 | 1.75290 | 52.137 | - | - | **1.754** |
| 220 | 1.23949 | 76.846 | **75.95** | 1.253 | **1.241** |
| 311 | 1.05704 | 93.560 | - | - | **1.030** |
| 222 | 1.01204 | 99.833 | - | *1.013* | **1.018** |
| | | | | | |
| a | 3.50580 | | | *3.507* | *3.506* |

| Cu(hkl) | *fcc* Cu ref [Å] | Cu ref (XRD) [°] | Cu exp (XRD) [°] | Cu exp (XRD) [Å] | Cu exp (TEM) [Å] |
|---|---|---|---|---|---|
| 111 | 2.09074 | 43.238 | **43.35** | 2.087 | **2.088** |
| 200 | 1.81063 | 50.356 | **50.54** | 1.856 | **1.809** |
| 220 | 1.28031 | 73.976 | - | - | -- |
| 311 | 1.09185 | 89.739 | - | - | -- |
| 222 | 1.04537 | 94.977 | - | - | **1.055** |
| | | | | | |
| a | 3.62126 | | | *3.615* | *3.618* |

"-" indicates that the corresponding 2θ peak for the crystal plane is out of the used XRD scan range, and "--" means that the crystal plane was not observed in TEM.

## 3. HAADF-STEM tomography reconstruction of NW morphology

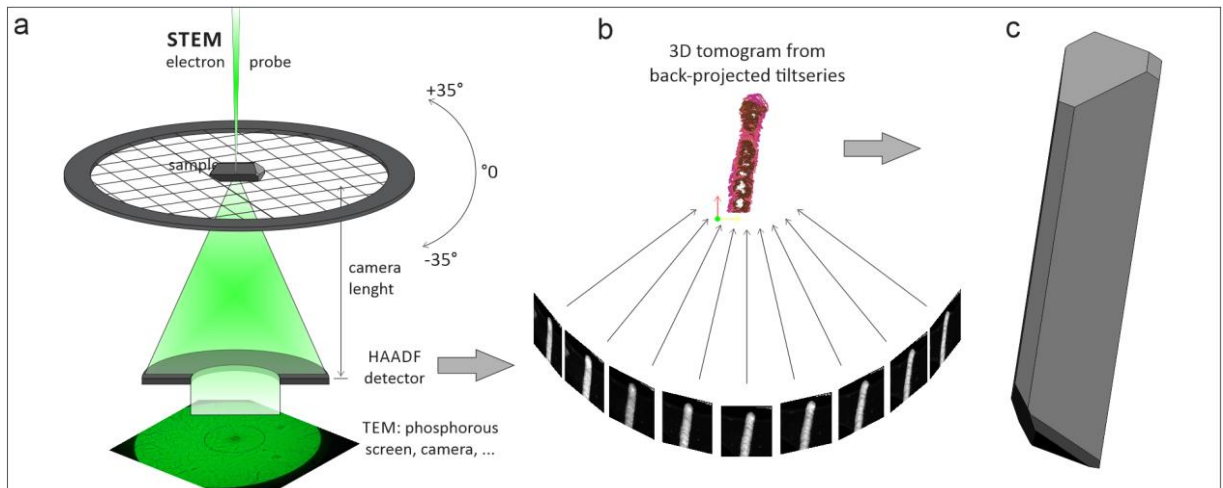

**Figure S3: HAADF-STEM tilt series used for reconstruction of the NW morphology.** (a) General schematics of the HAADF-STEM tilting experiment, with signal routes and reconstruction process. (b) The tilt series was constructed from 2° tilt step micrographs, in angular range of -30° to +40°, whereas at 0° the NW was in the 01-1 zone axis. (c) Final surface morphology was reconstructed from the NW outlines.

# 4. TEM; Lorentz electron microscopy used for visualisation of the magnetic domains in the single NW

Lorentz transmission electron microscopy (TEM) has been used to study magnetic domain structures (topological domains), as it is able to visualise the nanometric topological magnetic configurations such as vortices, bubbles, and skyrmions. Here, the Ni segments are confirmed to have a single-magnetic-domain structure.

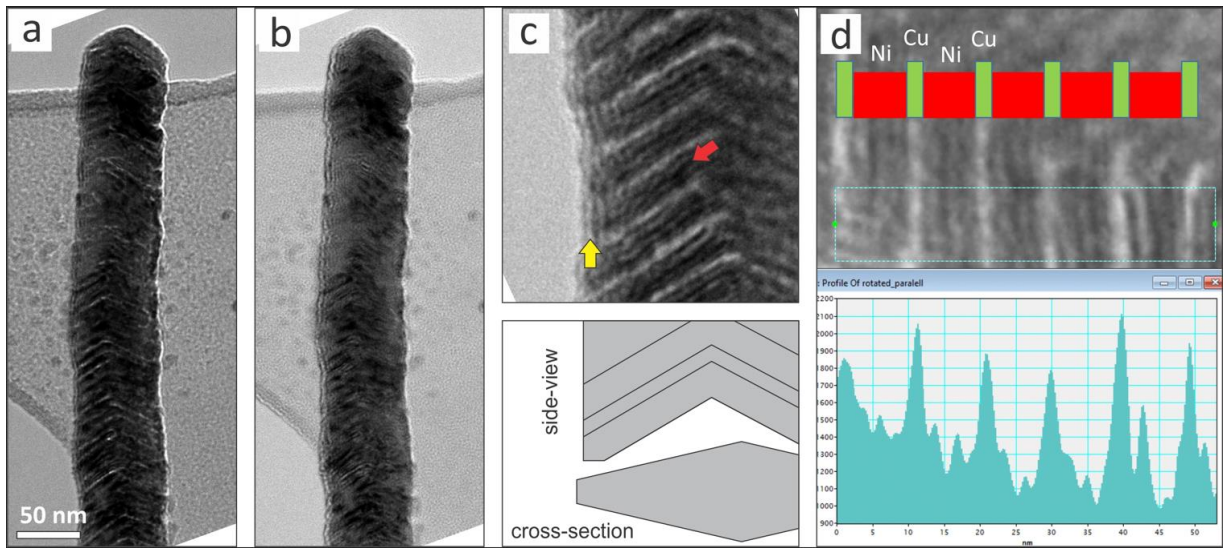

**Figure S4**: Fresnel-type TEM Lorentz microscopy; (a) overfocus and (b) underfocus. (c) Detail of the NW with contrast alteration due interaction of the primary electron beam with the magnetic domain of the Ni segment (red arrow). Due to the small size of the Ni, the whole layer present a single magnetic domain. Additional contrast on the side of the NW, in the thinnest edge (yellow arrow) is an effect of the morphology (thickness fringes) and shows a limitation of Lorentz microscopy when applied to faceted particulates, resulting in artefacts originating from Fresnel fringes easily mistaken for the signals from magnetic interactions. (d) Intensity profile of the contrast formation in Lorentz microscopy (in defocus and underfocus).

## 5. Wulff shapes, surface energy and work function

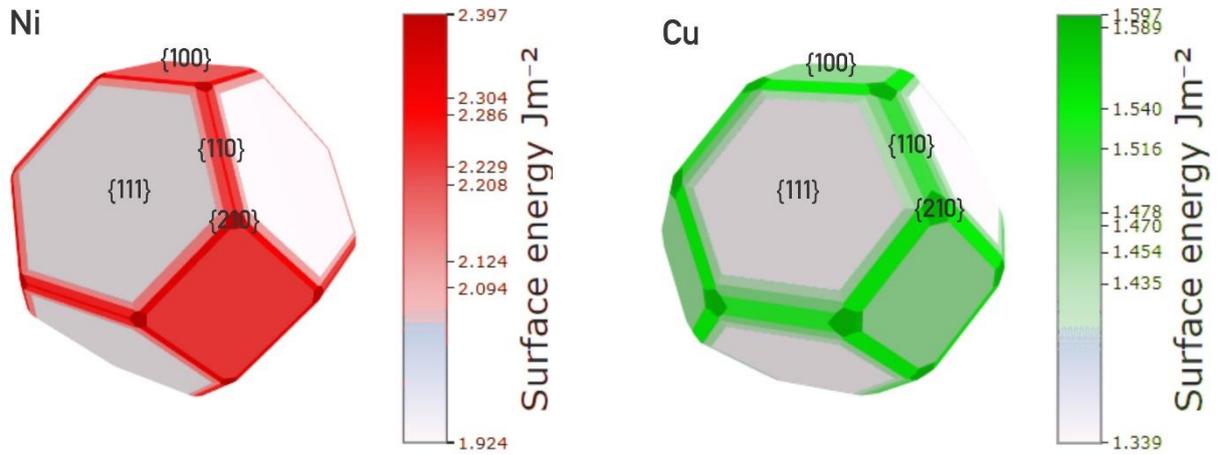

**Figure S5**: surface energies of Ni (red, a = 3.48 Å) and Cu (green, a = 3.58 Å), both for cubic Fm-3m symmetry. For both materials, the {111} facets have the lowest surface energy and the highest work function. After: Tran, R., *et al. Surface energies of elemental crystals*. Sci Data 3, 160080 (2016). DOI: 10.1038/sdata.2016.80

**Suppoting data Table S2:** Calculated surface energies (γ) and work function (Φ) for possible Ni and Cu facets. With * we marked the facets experimentally observed on the Ni/Cu nanowires.

| Ni(hkl) Miller indices | surface energy (γ) [J/m²] | surface energy (γ) [eV/Å²] | work function (Φ) [eV] | Cu(hkl) Miller indices | surface energy (γ) [J/m²] | surface energy (γ) [eV/Å²] | work function (Φ) [eV] |
|---|---|---|---|---|---|---|---|
| *(111) | 1.92 | 0.120 | 5.101 | *(111) | 1.34 | 0.084 | 4.711 |
| (332) | 2.09 | 0.131 | 4.858 | (332) | 1.43 | 0.090 | 4.526 |
| (322) | 2.12 | 0.133 | 4.789 | (322) | 1.45 | 0.091 | 4.380 |
| (221) | 2.17 | 0.136 | 4.654 | *(100) | 1.47 | 0.092 | 4.474 |
| *(100) | 2.21 | 0.138 | 4.954 | (221) | 1.48 | 0.092 | 4.356 |
| (331) | 2.23 | 0.139 | 4.515 | (331) | 1.52 | 0.095 | 4.306 |
| (211) | 2.24 | 0.140 | 4.656 | (311) | 1.54 | 0.096 | 4.278 |
| *(110) | 2.29 | 0.143 | 4.434 | *(110) | 1.56 | 0.097 | 4.195 |
| (311) | 2.30 | 0.144 | 4.538 | (321) | 1.58 | 0.099 | 4.217 |
| (321) | 2.32 | 0.145 | 4.473 | (310) | 1.59 | 0.099 | 4.171 |
| (320) | 2.39 | 0.149 | 4.507 | (210) | 1.60 | 0.100 | 4.176 |
| (310) | 2.40 | 0.150 | 4.567 | (320) | 1.62 | 0.101 | 4.267 |
| (210) | 2.40 | 0.150 | 4.414 | (211) | 1.63 | 0.102 | 4.224 |
| Weighted surf. e./ w.f. | 2.04 | 0.127 | 4.98 | Weighted surf. e./ w.f. | 1.42 | 0.089 | 4.53 |

**Supporting data Table S3:** Ni-Cu epitaxy cell mismatch calculation. Data used: **Ni**; Fm-3m, a = 3.51 A, V = 43.09 A$^3$, **Cu**; Fm-3m, a = 3.62 A, V = 47.49 A$^3$

| Substrate | film | MCIA [A$^2$] |
|---|---|---|
| Ni<100> | Cu<100> | 13.08 |
| Ni<110> | Cu<110> | 18.50 |
| Ni<111> | Cu<111> | 22.65 |
| Cu<100> | Ni<100> | 110.76 |
| Cu<111> | Ni<100> | 184.59 |
| Cu<110> | Ni<110> | 156.63 |

Minimal co-incident area (MCIA) is calculated by method: A. Zur; T. C. McGill. *Lattice match. An application to heteroepitaxy.* J. Appl. Phys. 55, 378–386 (1984) https://doi.org/10.1063/1.333084.

# 6. Crystal structure simulations and the Ni – Cu interface

**Ni-Cu shared plane: {010}; not observed in our system**

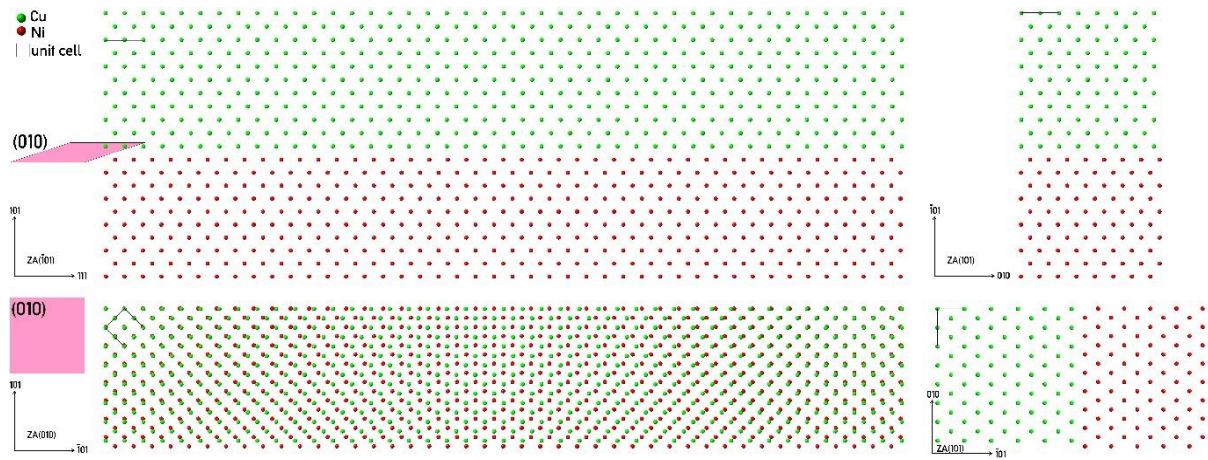

**Ni-Cu shared plane: {111}; observed in our system**

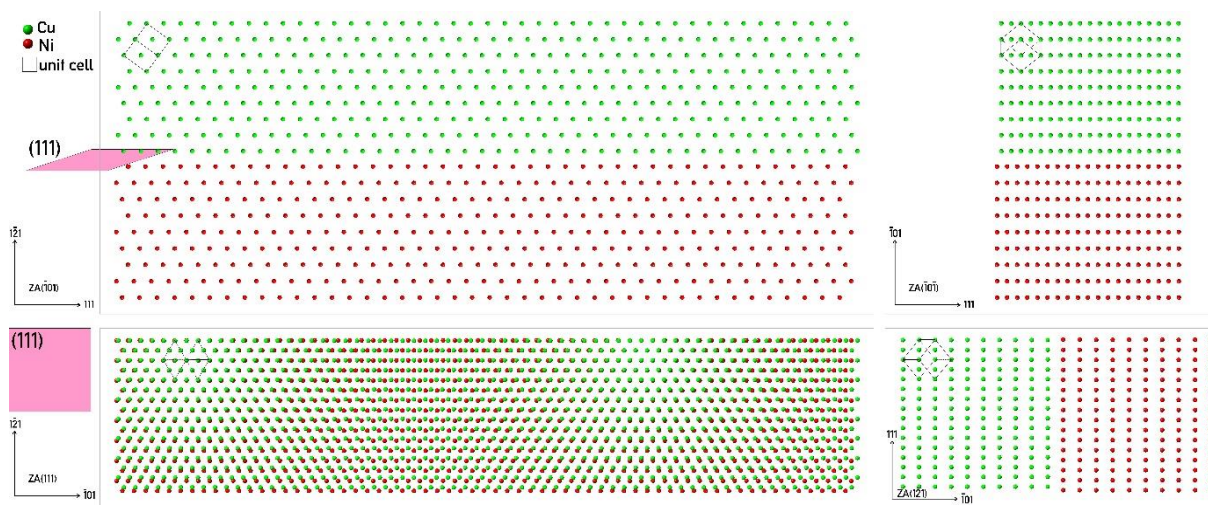

**Figure S6**: Crystal structure simulations and the Ni – Cu interface; both model structures are first visualised perpendicular (edge-on), and then parallel to the interface, i.e. 90 ° rotation around shared plane. **Example 1: *Ni-Cu (010) shared plane.*** An example of stable epitaxial simple cube-on-cube Ni-Cu interface, which was not experimentally observed in our system; the shared plane on the interface is (010). **Example 2: *Ni-Cu (111) shared plane.*** The simulated Cu-Ni interface with shared (111) plane, as observed in our system. The mismatch between Cu atoms (2.556 A) and Ni atoms (2.492 A) is 2.5 %, resulting in an extra Ni column every 40 atoms.